\documentclass{aa}

\usepackage{txfonts}
\usepackage{color}
\usepackage[normalem]{ulem}
\usepackage{xcolor}
\usepackage[normalem]{ulem}

\begin{document}

\title{Benford's law in the {\it Gaia} universe}

\author{
Jurjen de Jong\inst{\ref{inst:ESTEC},\ref{inst:Leuven},\ref{inst:matrixian}} \and Jos de Bruijne\inst{\ref{inst:ESTEC}} \and Joris De Ridder\inst{\ref{inst:Leuven}}
}

\institute{
Science Support Office, Directorate of Science, European Space Research and Technology Centre (ESA/ESTEC), Keplerlaan 1, 2201 AZ Noordwijk, The Netherlands\relax\label{inst:ESTEC}
\and Institute of Astronomy, KU Leuven, Celestijnenlaan 200D, 3001 Leuven, Belgium\relax\label{inst:Leuven}
\and Matrixian Group, Transformatorweg 104, 1014 AK Amsterdam, The Netherlands\relax\label{inst:matrixian}
}

\date{Received XXX 2019 / Accepted XXX 2020}

\begin{abstract}
    {
        Benford's law states that for scale- and base-invariant data sets covering a wide dynamic range, 
        the distribution of the first significant digit is biased towards low values.
        This has been shown to be true for wildly different datasets, including financial, geographical, and atomic data.
        In astronomy, earlier work showed that Benford's law also holds for distances estimated as the inverse of
         parallaxes from the ESA {\it Hipparcos} mission.    
    }
    {
        We investigate whether Benford's law still holds for the 1.3~billion parallaxes contained 
        in the second data release of {\it Gaia} ({\it Gaia} DR2). In contrast to previous work, we also include negative parallaxes. We examine whether distance estimates computed using a Bayesian approach
        instead of parallax inversion still follow Benford's law. Lastly, we investigate the use of Benford's law as a validation tool for the zero-point of the {\it Gaia} parallaxes.
    }
    {
        We computed histograms of the observed most significant digit of the parallaxes and distances, and compared them with the predicted values from Benford's law, as well as with 
        theoretically expected histograms. The latter were derived from a simulated {\it Gaia} catalogue based on the 
        Besan\c{c}on galaxy model.  
    }
    {
        The observed parallaxes in {\it Gaia} DR2 indeed follow Benford's law. 
        Distances computed with the Bayesian approach of \citet{2018AJ....156...58B} no longer follow Benford's law,
        although low-value ciphers are still favoured for the most significant digit. The prior that is used 
        has a significant effect on the digit distribution. Using the simulated
        \emph{Gaia} universe model snapshot, we demonstrate that the true distances underlying the {\it Gaia} catalogue are not 
        expected to follow Benford's law, 
        essentially because the interplay between the luminosity function of the Milky Way 
        and the mission selection function results in a bi-modal distance distribution, corresponding to nearby dwarfs in the 
        Galactic disc and distant giants in the Galactic bulge. In conclusion, {\it Gaia} DR2 parallaxes only follow Benford’s Law as a result of observational errors.
        Finally, we show that a zero-point
        offset of the parallaxes derived by optimising the fit between the observed most-significant digit frequencies and 
        Benford's law leads to a value that is inconsistent with the value that is derived from quasars. The underlying reason
        is that such a fit primarily corrects for the difference in the number of positive and negative parallaxes, and can thus not be used to obtain a reliable zero-point.
    }
    {}
    \maketitle
\end{abstract}

\section{Introduction}\label{sec:introduction}

Benford's law, sometimes referred to as the law of anomalous numbers or the significant-digit law, was put forward by Simon Newcomb in 1881 and later made famous by Frank Benford \citep{1881AmJM....4...39N,10.2307/984802}. The law states that the frequency distribution of the first significant digit of data sets representing (natural) phenomena covering a wide dynamic range such as terrestrial river lengths and mountain heights is non-uniform, with a strong preference for low numbers. As an example, Benford's law states that digit 1 appears as the leading significant digit 30.1\% of the time, while digit 9 occurs as first significant digit for only 4.6\% of data points taken from data sets that adhere to Benford's law. Although Benford's law has been known for more than a century and has received significant attention in a wide range of fields covering natural and (socio-)economic sciences \cite[e.g.][]{BL2015}, a statistical derivation was only published fairly recently, showing that Benford's law is the consequence of a central-limit-theorem-like theorem for significant digits \citep{MR1421567}.

\cite{2014JApA...35..639A} investigated the presence of Benford's law in the Universe and demonstrated that the $\sim$118,000 stellar parallaxes from the ESA {\it Hipparcos} astrometry satellite \citep{1997ESASP1200.....E}, converted into distances by inversion, follow Benford's law. In this paper, we extend this work and present an investigation into the intriguing question whether the $\sim$1.3 billion parallaxes and the associated Bayesian-inferred distances that are contained in the second data release of the {\it Hipparcos} successor mission, {\it Gaia} \citep{2016A&A...595A...1G,2018A&A...616A...1G}, follow Benford's law as well. Moreover, we investigate the prospects of using Benford's law as tool for validating the {\it Gaia} parallaxes. The idea of using Benford's law as a tool for anomaly detection is not new: \cite{Nigrini} described that Benford's law was used to detect fraud in income-tax declarations. We adapt this idea and investigate the effect of the parallax zero-point offset that is known to be present in the {\it Gaia} DR2 parallax data set \citep{2018A&A...616A...2L} with the aim to determine whether Benford's law can be used to derive the value of the offset.

This paper is organised as follows. Section~\ref{sec:BL} presents a short overview of Benford's law. Section~\ref{sec:hipparcos} summarises and discusses the {\it Hipparcos}-based study by \cite{2014JApA...35..639A} that inspired this work. Section~\ref{sec:gaia_dr2} presents our work, which is based on {\it Gaia} DR2. The effect of the parallax zero-point is discussed in Section~\ref{sec:zero_point}, and a discussion and conclusions can be found in Section~\ref{sec:conclusions}.

\section{Benford's law}\label{sec:BL}

Benford's law is an empirical, mathematical law that gives the probabilities of occurrence of the first, second, third, and higher significant digits of numbers in a data set. In this paper, we limit ourselves to the first significant digit. We also investigated the second and third significant digits, but this did not yield additional insights into this study.

Every number $X \in \mathbb{R}_{> 0}$ can be written in scientific notation as $X = x \cdot 10^{m}$, where $1 \leq x < 10$ and $x \in \mathbb{R}_{> 0}$, $m \in \mathbb{Z}$. The quantity $x$ is called the \textit{\textup{significand}}. The first significant digit is therefore also the first digit of the significand. This formulation allows us to define the first-significant digit operator $D_{1}$ on number $X$ with a floor function:
\begin{equation}
D_{1}X = \lfloor{x}\rfloor{}.\label{eq:D1}
\end{equation}

According to Benford's law, the probability for a first significant digit $d = 1, \ldots, 9$ to occur is
\begin{equation}
P(D_{1}X=d) = \log_{10}\left(1+\frac{1}{d}\right).\label{eq:BL}
\end{equation}
Benford's law states that the probability of occurrence of $1$ as first significant digit ($d=1$) equals $P(D_{1}X=1) = 0.301$. This probability decreases monotonically with higher numbers $d$, with $P(D_{1}X=2) = 0.176$, $P(D_{1}X=3) = 0.125$, down to $P(D_{1}X=9) = 0.046$ for $d=9$ as first significant digit.

\begin{figure}[t]
    \begin{center}
    \includegraphics[width=1.0\columnwidth]{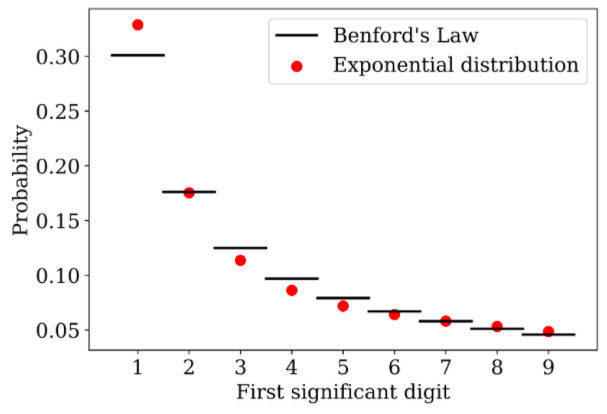}
    \end{center}
 \caption{Comparison of the frequency of occurrence of all possible values of the first significant digit ($d=1,\ldots,9$) between one million randomly drawn numbers from an exponential distribution ($e^{-X}$; red circles) and Benford's law (black, horizontal bars).}
    \label{fig:exponential_distribution}
\end{figure}

\begin{figure}[t]
  \begin{center}
    \includegraphics[width=1.0\columnwidth]{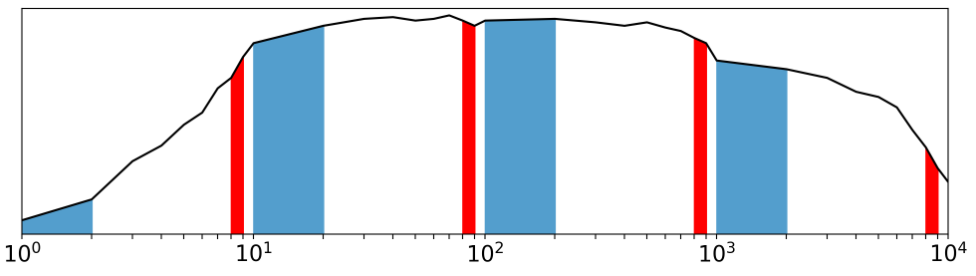}
  \end{center}
    \caption{Schematic example of a probability distribution of a variable that covers several orders of magnitude and that is fairly uniformly distributed on a logarithmic scale. The sum of the area of the blue bins is the relative probability that the first significant digit equals 1 ($d=1$), while the sum of the area of the red bins is the relative probability that the first significant digit equals 8 ($d=8$). Because the distribution is fairly uniform, i.e. the bin heights are roughly the same, the cumulative red and blue areas are foremost proportional to the fixed widths of the red and blue bins, respectively, such that numbers randomly drawn from this distribution will approximate Benford's law.
    \label{fig:logarithmic_axis}}
\end{figure}

Figure~\ref{fig:exponential_distribution} shows the first significant digit of randomly drawn numbers from an exponential distribution ($e^{-X}$) versus Benford's law. This shows that the exponential distributed data approaches Benford's law.

Benford's law is an empirical law. This means that there is no solid proof to show that a data set agrees with Benford's law. Nonetheless, the following conditions make it very likely that a data set follows Benford's law:
\begin{enumerate}

\item The data shall be non-truncated and rather uniformly distributed over several orders of magnitude. This can be understood through Eq.~(\ref{eq:BL}), which shows that on a logarithmic scale, the probability $P(D_{1}X=d)$ is proportional to the space between $d$ and $d+1$. In other words: Benford's law results naturally if the mantissae (the fractional part)  of the logarithms of the numbers are uniformly distributed. For example, the mantissa of $\log_{10}(2\cdot 10^{m}) \approx m + 0.30103$ for $m \in \mathbb{Z}$, equals $0.30103$. This is graphically illustrated in Figure~\ref{fig:logarithmic_axis}, in which we intuitively show that a distribution close to a uniform logarithmic distribution should obey Benford's law. In particular, the red bins with $d=1$ occupy $\sim$30\% of the axis, compared to the $\sim$5\% length of the blue bins that contain numbers where $d=8$. Even though the distribution is not perfectly uniform (the heights of the bins vary), the cumulative areas of all red and all blue bins are determined more by the (fixed) widths of the bins than by their heights, such that Benford's law is approximated when adding (i.e., averaging over) several orders of magnitude.
\item From the previous point, it follows intuitively that if a data set follows Benford's law, it must be scale invariant (see Appendix~\ref{subsec:scale_invariance}). In particular, a change of units, for instance from parsec to light year when stellar distances are considered, should not (significantly) change the probabilities of occurrence of the first significant digit. 
This can be understood by looking at Figure~\ref{fig:logarithmic_axis} and by considering a uniform logarithmic distribution for which the logarithmic property $\log{Cx} = \log{C} + \log{x}$ holds for variable $x \in \mathbb{R}_{> 0}$ and constant (scaling factor) $C \in \mathbb{R}_{> 0}$.
\item \citet{MR1421567} demonstrated that scale-invariance implies base-invariance (but not conversely). It therefore follows that if a data set follows Benford's law, it must be base invariant (see Appendix~\ref{subsec:base_invariance}). In particular, a change of base, for instance from base 10 as used in Eq.~(\ref{eq:BL}) to base 6, should not (significantly) change the probabilities of occurrence of the first significant digit in comparison to Benford's law.
\end{enumerate}

For reasons explained in Appendix~\ref{sec:stats}, we used a simple Euclidean distance to quantify how well the distribution of the first significant digit of {\it Gaia} data is described by Benford's law. Although this sample-size-independent metric is not a formal test statistic, with associated statistical power, this limitation is acceptable in this work because we only use the Euclidean distance as a relative measure (see Appendix~\ref{sec:stats} for an extensive discussion).

\begin{figure*}[t]
  \begin{center}
    \includegraphics[width=1.0\textwidth]{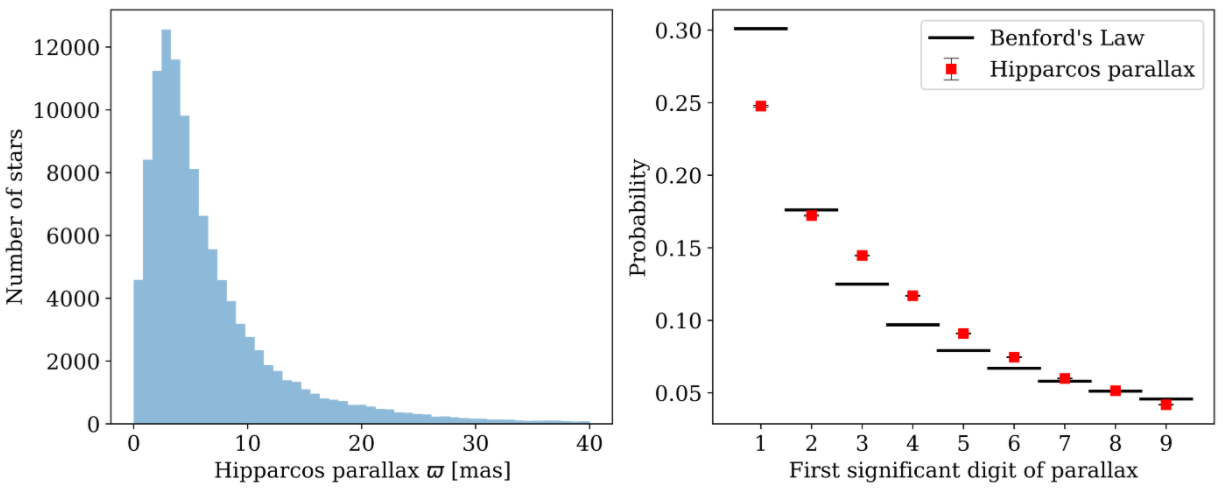}
    \caption{{\it Left:} {\it Hipparcos} parallax histogram for all $113\,942$ stars from \cite{2007ASSL..350.....V} with $\varpi > 0$~mas (1728 objects fall outside the plotted range). {\it Right:} Distribution of the first significant digit of the {\it Hipparcos} parallaxes together with the theoretical prediction of Benford's law. The data have vertical error bars to reflect Poisson statistics, but the error bars are much smaller than the symbol sizes.}
    \label{fig:hipparcos_parallax}
  \end{center}
\end{figure*}

\begin{figure*}[t]
  \begin{center}
    \includegraphics[width=1.0\textwidth]{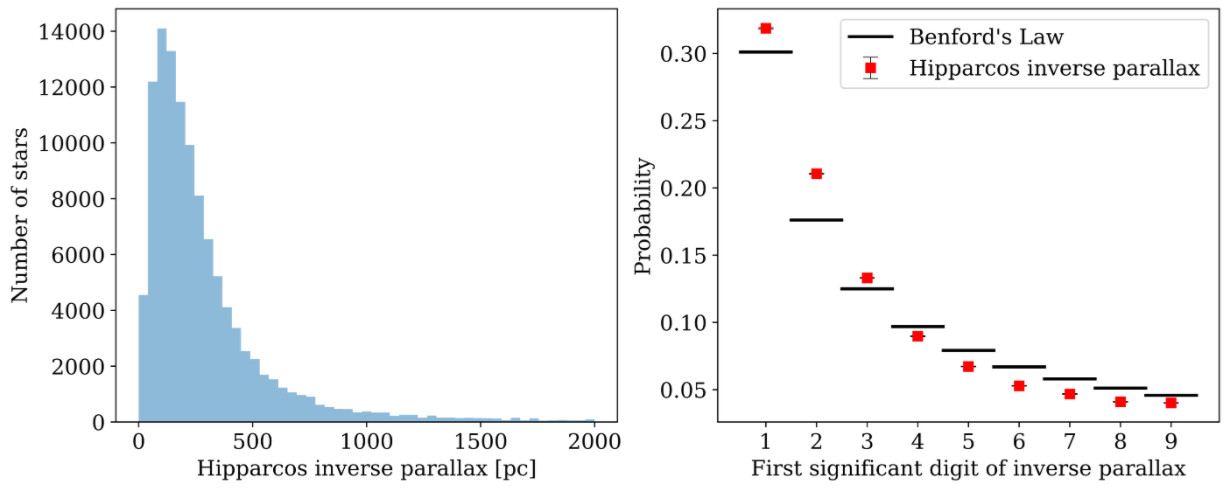}
    \caption{{\it Left:} {\it Hipparcos} inverse-parallax histogram for all $113\,942$ stars from \cite{2007ASSL..350.....V} with $\varpi > 0$~mas (3803 objects fall outside the plotted range). {\it Right:} Distribution of the first significant digit of the inverse parallaxes, referred to as ``distances'' by \cite{2014JApA...35..639A}, together with the theoretical prediction of Benford's law; compare with Figure~3(b) in \cite{2014JApA...35..639A}. The data have vertical error bars to reflect Poisson statistics, but the error bars are much smaller than the symbol sizes.}
    \label{fig:hipparcos_distance}
  \end{center}
\end{figure*}

\begin{figure*}[t]
  \begin{center}
    \includegraphics[width=1.0\textwidth]{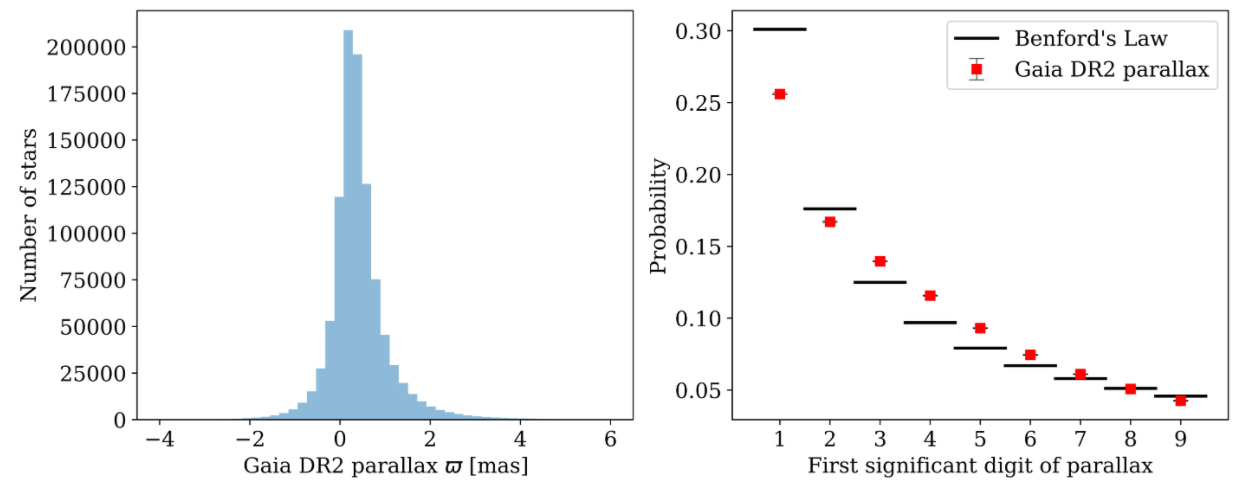}
    \caption{{\it Left:} {\it Gaia} DR2 parallax histogram (for a random sample of one million objects); 2305 objects fall outside the plotted range. {\it Right:} Distribution of the first significant digit of the absolute value of the {\it Gaia} DR2 parallaxes together with the theoretical prediction of Benford's law. The data have vertical error bars to reflect Poisson statistics, but the error bars are much smaller than the symbol sizes.}
    \label{fig:gaia_dr2_parallax}
  \end{center}
\end{figure*}

\section{{\it Hipparcos}}\label{sec:hipparcos}

\cite{2014JApA...35..639A} presented an assessment of Benford's law in relation to stellar distances. Unfortunately, the authors provide very limited information and just state that they used the distances from the HYG [{\it Hipparcos}-Yale-Gliese] database, which includes $115\,256$ stars with distances reaching up to 14 kpc. Based on various tests we conducted, trying to reproduce the results presented by \citeauthor{2014JApA...35..639A}, we conclude the following:
\begin{itemize}
\item The assessment of \citeauthor{2014JApA...35..639A} has been based on the HYG 2.0 database (September 2011), which consists of the original {\it Hipparcos} catalogue \citep{1997ESASP1200.....E} that contains $118\,218$ entries, $117\,955$ of which have five-parameter astrometry (position, parallax, and proper motion), merged with the fifth edition of the Yale Bright Star Catalog that contains 9110 stars  and the third edition of the Gliese Catalog of Nearby Stars that contains 3803 stars. We verified that changing the data set to the latest version of the catalogue, HYG 3.0 (November 2014), which is based on the new reduction of the {\it Hipparcos} data by \cite{2007ASSL..350.....V}, does not dramatically change the results.
\item The assessment of \citeauthor{2014JApA...35..639A} has simply removed the small number of entries with non-positive parallax measurements. Presumably, this was done because negative parallaxes, which are a natural but possibly non-intuitive outcome of the astrometric measurement process underlying the {\it Hipparcos} and also the {\it Gaia} mission, cannot be directly translated into distance estimates (see the discussion in Section~\ref{sec:gaia_dr2}). Because the fraction of {\it Hipparcos} entries with zero or negative parallaxes is small (4245 and 4013~objects, corresponding to 3.6\% and 3.4\% for the data from \citeauthor{1997ESASP1200.....E} or \citeauthor{2007ASSL..350.....V}, respectively), excluding or including them does not fundamentally change the statistics.
\item Distances have been estimated by \citeauthor{2014JApA...35..639A} from the parallax measurements through simple inversion. Whereas the {\it \textup{true}} parallax and distance of a star are inversely proportional to each other, the estimation of a distance from a {\it \textup{measured}} parallax, which has an associated uncertainty (and which can even be formally negative) requires care to avoid biases and to derive meaningful uncertainties \citep[see e.g.][]{2018A&A...616A...9L}. While for relative parallax errors below $\sim$10--20\% a distance estimation by parallax inversion is an acceptable approach, Bayesian methods are superior for a distance estimation for larger relative parallax errors (see the discussion in Section~\ref{subsec:gaia_dr2_distance}). In the case of the {\it Hipparcos} data sets, only 42\% and 51\% of the objects from \citeauthor{1997ESASP1200.....E} and \citeauthor{2007ASSL..350.....V}, respectively, have relative parallax errors below 20\%. In general,  the ``distances'' inferred by \citeauthor{2014JApA...35..639A} are therefore biased as well as unreliable.
\end{itemize}

Figure~\ref{fig:hipparcos_parallax} shows the histogram of the parallax measurements in the {\it Hipparcos} data set from \cite{2007ASSL..350.....V} and the associated first-significant-digit distribution. The latter resembles Benford's law, at least trend-wise, although we recognise that it is statistically speaking not an acceptable description. Nonetheless, the overabundance of low-number compared to high-number digits is striking and significant. 

Following Appendix~\ref{subsec:inverse_invariance}, the distribution of inverse parallaxes, that suggestively but incorrectly are referred to as ``distances'' by \cite{2014JApA...35..639A}, should also follow (the trends of) Benford's law. This is confirmed in Figure~\ref{fig:hipparcos_distance}. It shows that because small, positive parallaxes ($0 < \varpi \la 1$~mas) are abundant in the {\it Hipparcos} data set, many stars are placed (well) beyond 1 kpc in inverse parallax (``distance''). As a result, the inverse parallaxes (``distances'') span several orders of magnitude and resemble Benford’s Law. We conclude that the results obtained by \cite{2014JApA...35..639A} are reproducible but that their interpretation of inverse {\it Hipparcos} parallaxes as distances can be improved.

\begin{figure}[t]
  \begin{center}
    \includegraphics[width=0.9\columnwidth]{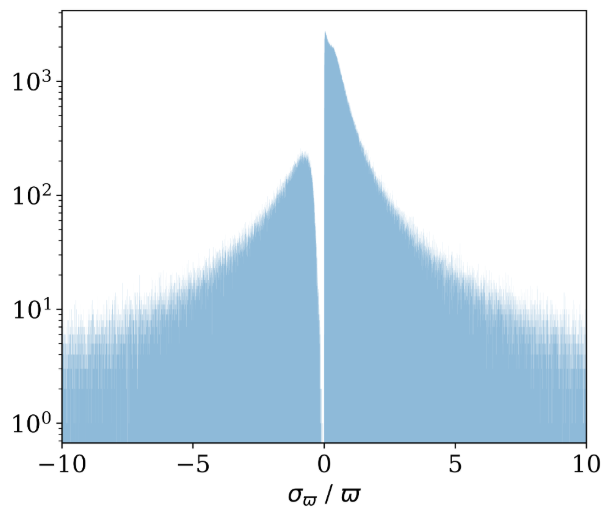}
    \caption{Central part of the distribution of relative parallax errors in the {\it Gaia} DR2 catalogue (the inverse of field {\tt parallax\_over\_error}).}
    \label{fig:gaia_dr2_relative_parallax_error}
  \end{center}
\end{figure}

\begin{figure*}[t]
  \begin{center}
    \includegraphics[width=1.0\textwidth]{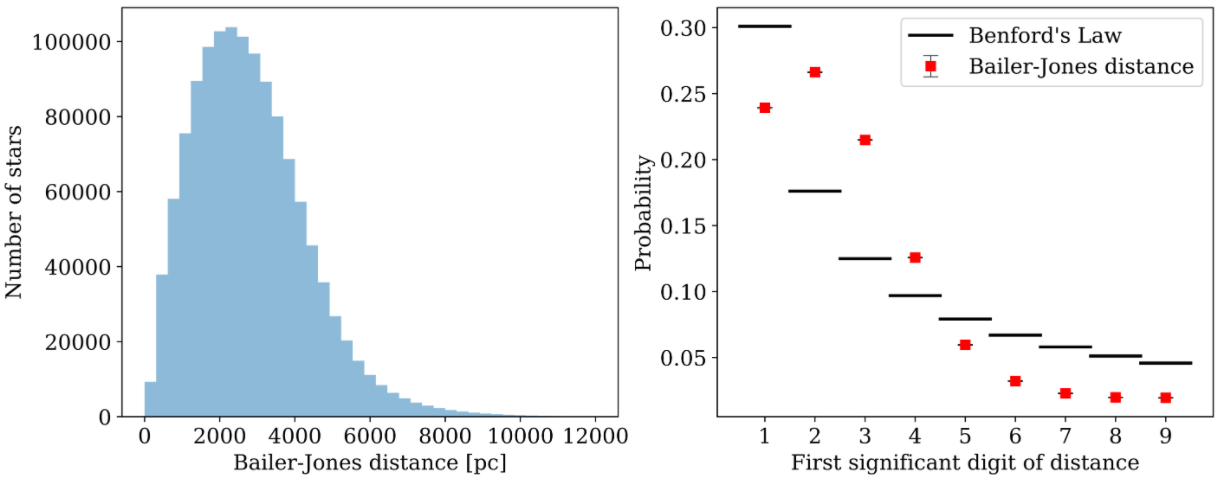}
    \caption{{\it Left:} Histogram of the {\it Gaia} DR2 Bayesian distance estimates from \cite{2018AJ....156...58B} for a random sample of one million objects (308 objects fall outside the plotted range). {\it Right}: Distribution of the first significant digit of the Bayesian distance estimates, together with the theoretical prediction of Benford's law. The data have vertical error bars to reflect Poisson statistics, but the error bars are much smaller than the symbol sizes.}
    \label{fig:gaia_dr2_distance}
  \end{center}
\end{figure*}

\begin{figure*}[t]
  \begin{center}
    \includegraphics[width=1.0\textwidth]{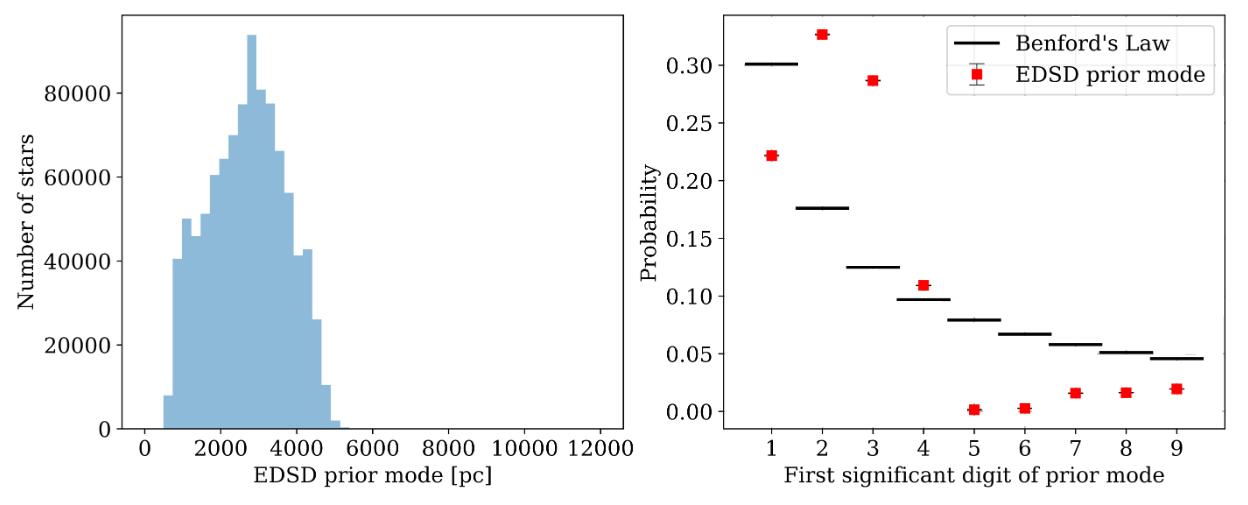}
    \caption{{\it Left:} Histogram of the mode of the EDSD prior (i.e. twice the exponential length scale $L = L(l,b)$) used by \cite{2018AJ....156...58B} for their Bayesian distance estimations of {\it Gaia} DR2 sources. {\it Right}: Distribution of the first significant digit of the EDSD prior mode, together with the theoretical prediction of Benford's law. The data have vertical error bars to reflect Poisson statistics, but the error bars are much smaller than the symbol sizes.}
    \label{fig:Bailer_Jones_mode}
  \end{center}
\end{figure*}

\begin{figure*}[t]
  \begin{center}
    \includegraphics[width=1.0\textwidth]{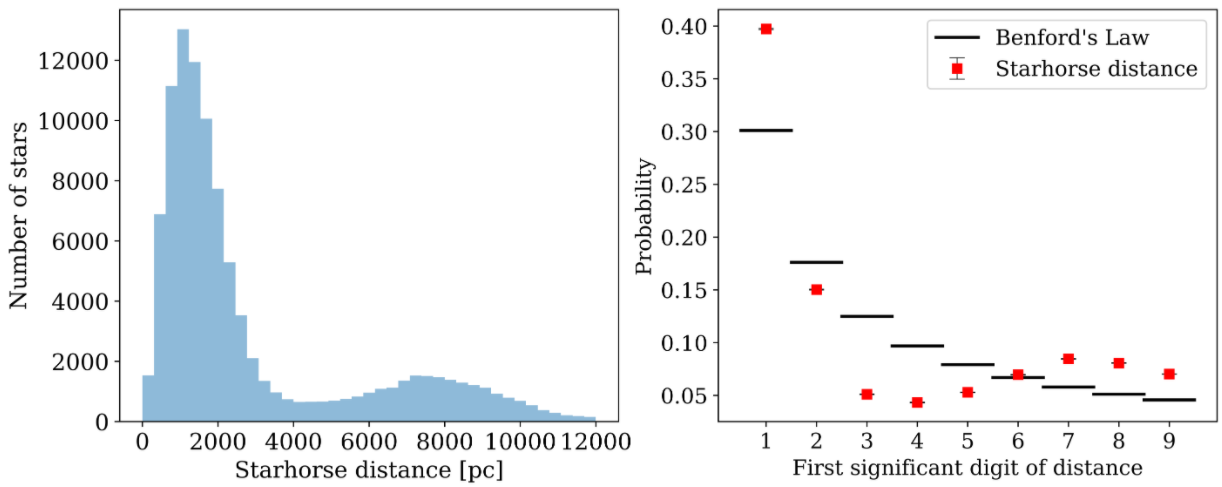}
    \caption{{\it Left:} Histogram of the subset of high-quality StarHorse distances from \cite{2019arXiv190411302A}. {\it Right}: Distribution of the first significant digit of these distances, together with the theoretical prediction of Benford's law. The data have vertical error bars to reflect Poisson statistics, but the error bars are much smaller than the symbol sizes.}
    \label{fig:StarHorse_distance}
  \end{center}
\end{figure*}

\begin{figure*}[t]
  \begin{center}
    \includegraphics[width=1.0\textwidth]{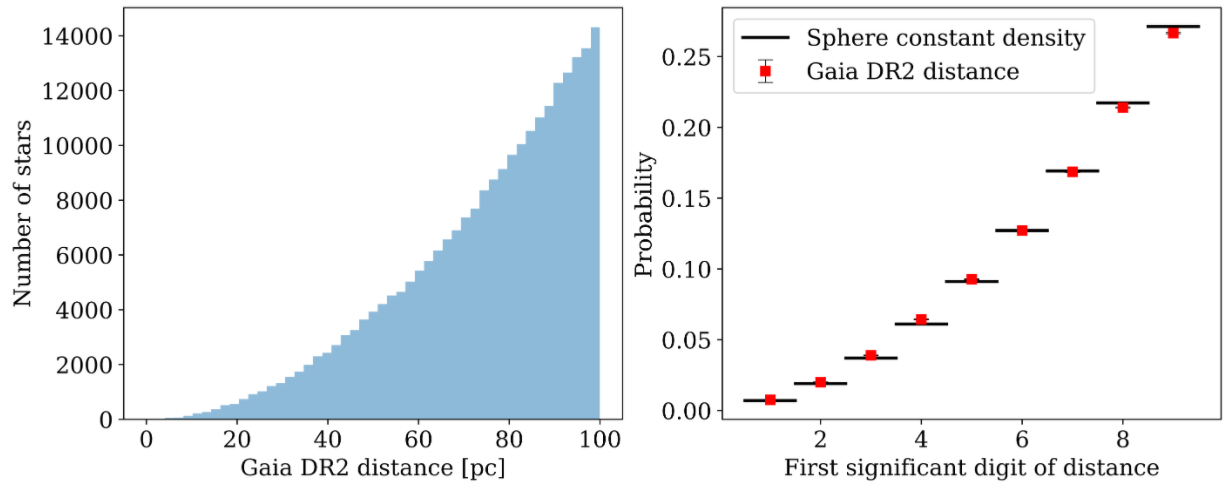}
    \caption{{\it Left:} Histogram of the {\it Gaia} DR2 Bayesian distance estimates from \cite{2018AJ....156...58B} for the sample of $243\,291$ stars within 100~pc from the Sun. {\it Right}: Distribution of the first significant digit of the {\it Gaia} DR2 Bayesian distance estimates displayed in the left panel, together with the theoretical prediction of a sample of stars with uniform, constant density. The data have vertical error bars to reflect Poisson statistics, but the error bars are much smaller than the symbol sizes.}
    \label{fig:gaia_dr2_distance_neighbourhood}
  \end{center}
\end{figure*}

\section{{\it Gaia} DR2}\label{sec:gaia_dr2}

The recent release of the {\it Gaia} DR2 catalogue \citep{2018A&A...616A...1G} offers a unique opportunity to make a study of Benford's law and stellar distances not based on $0.1$~million, but $1\,000+$ million objects. We discuss the parallaxes in Section~\ref{subsec:gaia_dr2_parallax} and the associated distance estimates in Sections~\ref{subsec:gaia_dr2_distance} and \ref{subsec:gaia_dr2_distance_neighbourhood}. We compare our findings with simulations in Section~\ref{subsec:simulations}.

\subsection{{\it Gaia} DR2 parallaxes}\label{subsec:gaia_dr2_parallax}

{\it Gaia} DR2 contains five-parameter astrometry (position, parallax, and proper motion) for $1\,331\,909\,727$ sources. One feature of the {\it Gaia} DR2 catalogue is the presence of spurious entries and non-reliable astrometry. Suspect data can be filtered out using quality filters recommended in \cite{2018A&A...616A...2L}, \cite{2018A&A...616A...4E}, and \cite{2018A&A...616A..17A}. Rather than filtering on the astrometric unit weight error (UWE), we filtered on the renormalised astrometric unit weight error (RUWE).\footnote{
  See the {\it Gaia} DR2 known issues website \url{https://www.cosmos.esa.int/web/gaia/dr2-known-issues} and Appendix~\ref{sec:filters} for details.
} In this study, we consistently applied the standard photometric excess factor filter published in April 2018 plus the revised astrometric quality filter published in August 2018 (see Appendix~\ref{sec:filters} for a detailed discussion). Another point worth mentioning is that in contrast to the {\it Hipparcos} case (Section~\ref{sec:hipparcos}), the percentage of non-positive parallaxes ({\it Gaia} DR2 does not contain objects with parallaxes that are exactly zero) in {\it Gaia} DR2 is substantial, at 26\% (Figure~\ref{fig:gaia_dr2_parallax}). Throughout this study, when we consider the first significant digit of {\it Gaia} DR2 parallaxes, we take the absolute value of the parallaxes first. The justification and implications of this choice are discussed in Appendix~\ref{sec:negative_parallaxes}.

Figure~\ref{fig:gaia_dr2_parallax} shows the distribution of the first significant digit of the {\it Gaia} DR2 parallaxes. In practice, to save computational resources, we use randomly selected subsets of the data throughout this paper, unless stated otherwise (see Appendix~\ref{sec:sample_size} for a detailed discussion). The first significant digit of the parallax sample follows Benford's law well. 

This is expected because at least two conditions identified in Section~\ref{sec:BL} are met. First of all, the histogram of the {\it Gaia} DR2 parallaxes shows that the distributions cover four orders of magnitude. Secondly, the distribution of the first significant digit of the parallaxes is not sensitive to scaling. That is, we verified that multiplying all parallaxes with a constant $C$ does not fundamentally change the distribution of the first significant digit: for instance, the probabilities for digit $d=1$ do not change by more than 5\%\  points by scaling the data with any factor $C$ in the range 1--10 (see also Appendix~\ref{sec:stats}). 
We postpone the discussion of why the first significant digit of the parallaxes follows Benford's law so closely to Section~\ref{subsec:simulations}.

\subsection{{\it Gaia} DR2 distance estimates}\label{subsec:gaia_dr2_distance}

As we reported in Section~\ref{sec:hipparcos}, astrometry missions such as {\it Hipparcos} and {\it Gaia} do not measure stellar distances but parallaxes. These measurements are noisy such that as a result of the non-linear relation between (true) parallax and (true) distance (${\rm distance} \propto {\rm parallax}^{-1}$), distances estimated as inverse parallaxes are fundamentally biased \citep[for a detailed disussion, see][]{2018A&A...616A...9L}. Whereas this bias is small and can hence be neglected, for small relative parallax errors (e.g. below $\sim$10--20\%), it becomes significant for less precise data. Figure~\ref{fig:gaia_dr2_relative_parallax_error} shows a histogram of the relative parallax error of the {\it Gaia} DR2 catalogue. It shows that only 9.9\% of the objects with positive parallax have $1 / {\tt parallax\_over\_error} < 0.2$, indicating that great care is needed in distance estimation.

As explained in \citet[][and references therein]{2018AJ....156...58B}, distance estimation from measured parallaxes is a classical inference problem that is ideally amenable to a Bayesian interpretation. this approach has the advantage that negative parallax measurements can also be physically interpreted and that meaningful uncertainties on distance estimates can be reconstructed. Using a distance prior based on an exponentially decreasing space density (EDSD) model, \cite{2018AJ....156...58B} presented Bayesian distance estimates for (nearly) all sources in {\it Gaia} DR2 that have a parallax measurement. Figure~\ref{fig:gaia_dr2_distance} compares the distribution of the first significant digit of these distance estimates to Benford's law. This time, a poor match can be noted: instead of 1 as the most frequent digit, digits 2 and 3 appear more frequently. Why the first significant digits of the Bailer-Jones distance estimates do not follow Benford's law is evident from their histogram: Most stars in {\it Gaia} DR2 are located at $\sim$2--3 kpc from the Sun (see also Figure~\ref{fig:gums_distance}). This is mostly explained by the EDSD prior adopted by \citeauthor{2018AJ....156...58B} in their Bayesian framework. For the (small) set of (nearby) stars with highly significant parallax measurements, the choice of this prior is irrelevant and the distance estimates are strongly constrained by the measured parallaxes themselves. For the (vast) majority of (distant) stars, however, the low-quality parallax measurements contribute little weight, and the distance estimates mostly reflect the choice of the prior. The EDSD prior has one free parameter, namely the exponential length scale $L,$ which can be tuned independently for each star. \cite{2018AJ....156...58B} opted to model this parameter as function of galactic coordinates $(\ell,b)$ based on a mock galaxy model. Because the EDSD prior has a single mode at $2 L$ and because $L$ ({\tt r\_len} in the data model) has been published along with the Bayesian distance estimates for each star, a prediction for the distribution of the first significant digit of the mode of the prior can be made accordingly. Figure~\ref{fig:Bailer_Jones_mode} shows this prediction, along with the actual distribution of the mode of the EDSD prior, for a random sample of one million stars. The digit distribution compares qualitatively well with that of the Bayesian distance estimates (cf.\ Figure~\ref{fig:gaia_dr2_distance}), with digit 2 appearing most frequently, followed by digits 1 and 3, followed by digit 4, and with digits 5--9 being practically absent. Quantitative differences between the digit distributions can be understood by comparing the left panels of Figures~\ref{fig:gaia_dr2_distance} and \ref{fig:Bailer_Jones_mode}. Whereas the distance distribution has a smooth, Rayleigh-type shape, extending out to $\sim$8~kpc, the prior mode distribution is noisy as a result of the extinction law applied in the mock galaxy model used by \cite{2018AJ....156...58B} and lacks signal below $\sim$700~pc and above $\sim$5~kpc. This finite range is a direct consequence of the way in which the length scale was defined by \citeauthor{2018AJ....156...58B}, who opted to compute it for $49\,152$ pixels on the sky as one-third of the median of the (true) distances to all the stars from the galaxy model in that pixel (and subsequently creating a smooth representation as function of Galatic coordinates $\ell,b$ by fitting a spherical harmonic model). This resulted in a lowest value of $L$ of 310~pc and a highest value of 3.143~kpc (such that the EDSD prior mode $2L$ can only take values between 620~pc and 6.286~kpc).

\cite{2019arXiv190411302A} published a set of $265\,637\,087$ photo-astrometric distance estimates obtained by combining {\it Gaia} DR2 parallaxes for stars with $G < 18$~mag with PanSTARRS-1, 2MASS, and AllWISE photometry based on the StarHorse code. The recommended quality filters SH\_GAIAFLAG="000" to select non-variable objects that meet the RUWE and photometric excess-factor filters from Appendix~\ref{sec:filters} and SH\_OUTFLAG="00000" to select high-quality StarHorse distance estimates leave $136\,606\,128$ objects. Figure~\ref{fig:StarHorse_distance} shows their distance histogram and the associated distribution of the first significant digit. The strong preference for digits 1 and 2, followed by digits 7 and 8, is explained by the bi-modality of the distance histogram, showing a strong peak of main-sequence dwarfs at $\sim$1.5~kpc and a secondary peak of (sub)giants in the Bulge around 7.5 kpc.

Our main conclusion of this section is that all available, large-volume, {\it Gaia}-based distance estimates prefer small leading digits. This fact, however, foremost reflects the structure of the Milky Way, combined with its luminosity function and extinction law, and the magnitude-limited nature of the {\it Gaia} survey.

\subsection{{\it Gaia} DR2 distance estimates in the solar neighbourhood}\label{subsec:gaia_dr2_distance_neighbourhood}

\begin{figure}[t]
  \begin{center}
    \includegraphics[width=1.0\columnwidth]{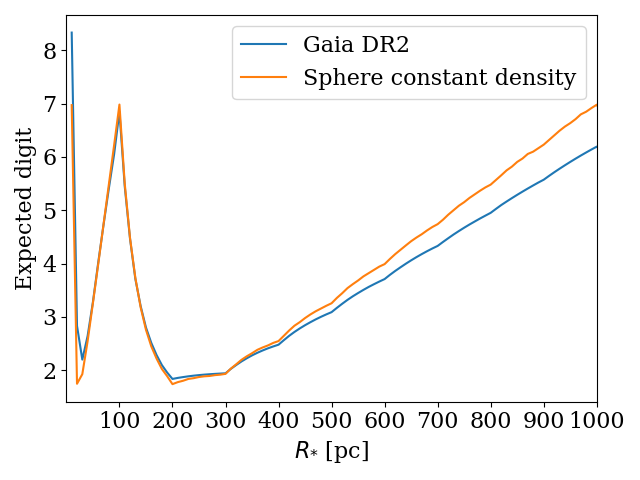}
    \caption{Comparison of the expectation value of the first significant digit of the distance distribution of (a) all {\it Gaia} DR2 stars with distances less than $R_\star$ and (b) a model of the solar neighbourhood in which stars have a uniform, constant density.}
    \label{fig:gaia_dr2_distance_neighbourhood_expectation}
  \end{center}
\end{figure}

It is to be expected that the distribution of the first significant digit of stellar distances is (much) less dependent on the prior for high-quality parallaxes, for which the distance estimates are strongly constrained by the parallax measurements themselves. To verify this, we show this distribution in Figure~\ref{fig:gaia_dr2_distance_neighbourhood} for the subset of $243\,291$ stars in {\it Gaia} DR2 that have distance estimates below $100$~pc and hence have typically small relative parallax errors (e.g. the mean and median values of parallax over error are 261 and 214, respectively). In this case, in contrast to the case of Benford's law, digit 1 appears least frequently and digit 9 appears most frequently. Assuming that stars in the solar neighbourhood are approximately uniformly distributed with a constant density, this can be understood because the volume between equidistant (thin) shells centred on the Sun increases with the cube of the shell radius (e.g. there are $[100^3-90^3]/[20^3-10^3] \sim 39$ times as many objects in the shell between 10 and 20~pc compared to the shell between 90 and 100~pc). As shown in Figure~\ref{fig:gaia_dr2_distance_neighbourhood}, the model of a constant-density solar neighbourhood is an almost perfect match with the data. 

In order to determine out to which distance this is true, Figure~\ref{fig:gaia_dr2_distance_neighbourhood_expectation} shows how the expectation value of the first significant digit of the distance distribution for all stars located within a sphere around the Sun with radius $R_\star$ varies with this radius and compares this expectation value with the one from the constant-density model. The (maybe initially surprising) variation in the expectation value between digits 2 and 7 with the distance limit of the sample (or the model) can be understood using the same argument as used before, linked to the cube dependence of the volume on distance. Fair agreement between data and model (difference $<$ 10\%) occurs up to $\sim$720~pc, which depending on extinction corresponds to early-M dwarfs for a faint limiting magnitude of 20.7~mag. To find a value of 720~pc is not surprising given the $\sim$300~pc vertical scale height of the thin disc \citep[e.g.][]{2016ARA&A..54..529B}. We conclude that whereas the distribution of the first significant digit of distances of the full Gaia DR2 catalogue is biased to digits 2 and 3 (Section~\ref{subsec:gaia_dr2_distance}) and mostly reflects the prior that has been used in the Bayesian estimation of the distances, local samples ($\la$720~pc) with high-quality parallaxes show a preference for a range of digits, with the most frequent digit depending on the limiting distance, which is fully compatible with a distribution of stars with uniform, constant density.

\subsection{Comparison with {\it Gaia} simulations}\label{subsec:simulations}

\begin{figure*}[t]
  \begin{center}
    \includegraphics[width=1.0\textwidth]{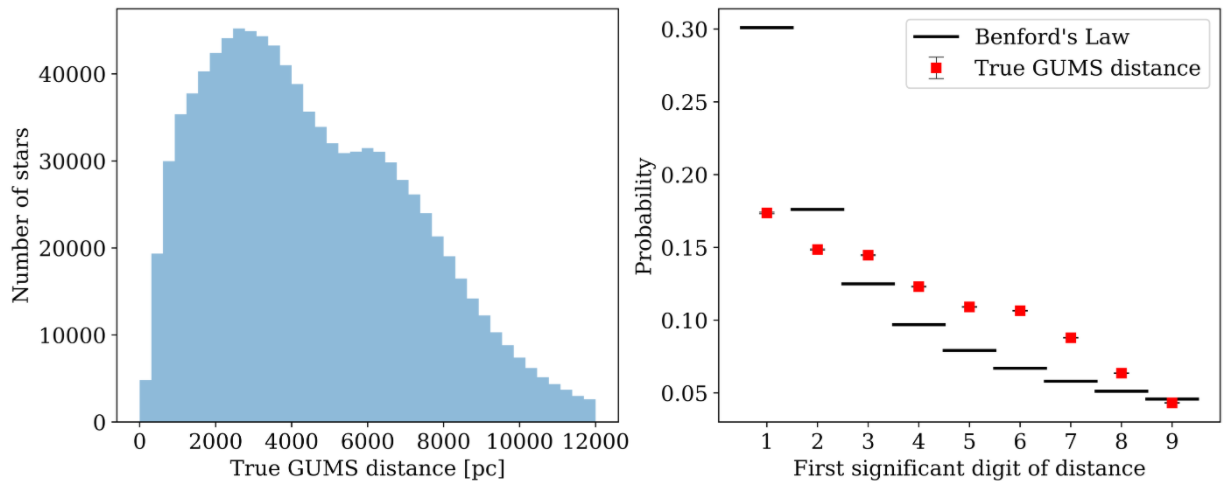}
    \caption{{\it Left:} Histogram of one million simulated true GUMS distances from \cite{2012A&A...543A.100R}; $19\,594$ objects fall outside the plotted range. {\it Right}: Distribution of their first significant digit together with the theoretical prediction of Benford's law. Figure~\ref{fig:gaia_dr2_distance} shows the same contents, but using the {\it Gaia} DR2 distance estimates from \cite{2018AJ....156...58B}. The data have vertical error bars to reflect Poisson statistics, but the error bars are much smaller than the symbol sizes.}
    \label{fig:gums_distance}
  \end{center}
\end{figure*}

\begin{figure*}[t]
  \begin{center}
    \includegraphics[width=1.0\textwidth]{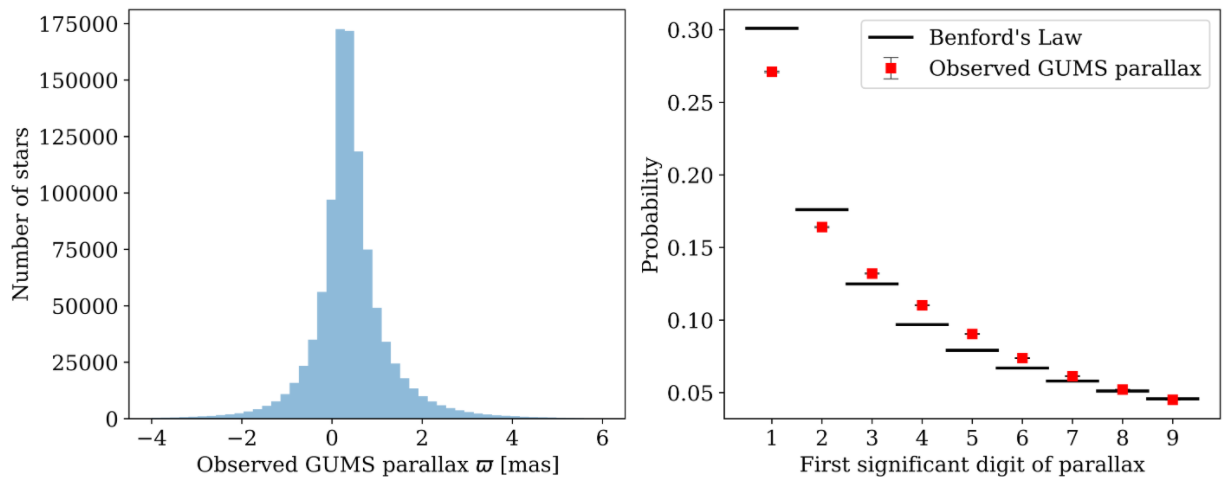}
    \caption{{\it Left:} Histogram of the simulated observed GUMS parallaxes from \cite{2014A&A...566A.119L}; $4879$ objects fall outside the plotted range. {\it Right:} Distribution of their first significant digit together with the theoretical prediction of Benford's law. The data have vertical error bars to reflect Poisson statistics, but the error bars are much smaller than the symbol sizes. Figure~\ref{fig:gaia_dr2_parallax} shows the same contents, but using the {\it Gaia} DR2 parallaxes.
    \label{fig:gog_parallax}}
  \end{center}
\end{figure*}

\cite{2012A&A...543A.100R} presented the {\it Gaia} universe model snapshot (GUMS). GUMS is a customised and extended incarnation of the Besan\c{c}on galaxy model, fine-tuned to a perfect {\it Gaia} spacecraft that makes error-less observations. GUMS represents a sophisticated, realistic, simulated catalogue of the Milky Way (plus other objects accessible to {\it Gaia,} such as asteroids and external galaxies) observable by {\it Gaia}, containing more than one~billion stars down to $G < 20$~mag. According to the Besan\c{c}on galaxy model, the Milky Way consists of an exponentially thin disc (67\% of the objects), an exponentially thick disc (22\% of the objects), a bulge (10\% of the objects), and a halo (1\% of the objects). Not surprisingly, given the luminosity function and magnitude-limited nature of the {\it Gaia} survey, the majority of the stars in GUMS are within a few kiloparsec from the Sun, with 69\% being main-sequence objects and 29\% being sub(giants). Figure~\ref{fig:gums_distance} shows the histogram of the true (i.e. noise-less, simulated) GUMS distances and the associated distribution of the first significant digits, which does not resemble Benford’s Law at all. The aforementioned bi-modality in distances, with a main disc peak around 2--3 kpc and a strong secondary bulge peak around 5-7 kpc, explains the relatively even occurrence of the numbers 1-7 as leading significant digit.

\cite{2014A&A...566A.119L} presented an observed version of the GUMS catalogue resulting from application of {\it Gaia}-specific error models that implement realistic observational errors that depend, as in reality, on object properties such as magnitude, on the {\it Gaia} instrument characteristics such as read-out noise, and on the number of observations made over the nominal five-year operational lifetime. The vast majority of the 523 million individual, single stars
are main-sequence dwarfs of spectral types F, G, K, and M (381 million, corresponding to 73\%) and (sub)giants of spectral type F, G, and K (133 million, corresponding to 25\%). Figure~\ref{fig:gog_parallax} shows the parallax distribution of this observed GUMS sample, together with the distribution of their first significant digit. When we compare Figure~\ref{fig:gog_parallax} with Figure~\ref{fig:gaia_dr2_parallax}, we recall that the first reflects a five-year {\it Gaia} mission and the second refers to {\it Gaia} DR2, which is based on 22 months of data (which implies that the formal uncertainties are to first order a factor $\sqrt{60/22} \approx 1.7$ larger). Nonetheless, the agreement between simulations and {\it Gaia} DR2 is striking.

When we compare Figure~\ref{fig:gums_distance} with Figure~\ref{fig:gog_parallax}, it is striking that the distribution of noise-free GUMS distances shows a larger departure from Benford's law than the distribution of noisy parallaxes simulated from them.
When the noise-free GUMS distances are inverted to noise-free parallaxes, the Euclidean distance of the first-significant-digit distribution with respect to Benford’s Law does not drastically change ($<$10\%), suggesting that not the inversion in itself, but the observational (parallax) errors are responsible for the improved match to Benford’s Law. This agrees with the inverse-invariance of first-significant-digit distributions that follow Benford's law (see Appendix \ref{subsec:inverse_invariance}). In order to investigate this further, we took the noise-free GUMS parallaxes and perturbed them with a parallax error that was randomly drawn (for each individual object) from a Gaussian distribution with zero mean and a fixed standard deviation that reflects the parallax standard error. The first significant digits of the associated distribution of noisy parallaxes can then be compared to Benford’s Law, and the agreement quantified through the Euclidean distance metric. When we repeated this for ever-increasing parallax standard errors ($\sigma_\varpi$), we found that the Euclidean distance with respect to Benford’s Law rapidly decreased by a factor $\sim$2, from 0.13 for $\sigma_\varpi = 0$~mas (noise-free GUMS parallaxes) to 0.06 for $\sigma_\varpi = 0.7$~mas (which is a typical parallax standard error for a representative faint star in {\it Gaia} DR2). In other words, the distribution of the first significant digits of the parallaxes approaches Benford’s Law more and more when the parallaxes are made more and more noisy. The reason for this significant reduction in Euclidean distance when observational parallax errors are added is shown in Figures~4 and 9 in \cite{2014A&A...566A.119L}. These figures show that for the vast majority of stars, the (end-of-life, so surely the {\it Gaia} DR2) parallax error is larger than the true parallax itself. Combined with the fact that more than 40\% of the GUMS stars have $d=1$ as leading significant digit for their true parallax, as dictated by the nature of the Milky Way and the {\it Gaia} survey, this causes a bell-shape of the distribution of observed parallaxes around a low mean parallax value (Figures~\ref{fig:gaia_dr2_parallax} and \ref{fig:gog_parallax}, left panels) such that the first significant digits nicely follow Benford’s Law (Figures~\ref{fig:gaia_dr2_parallax} and \ref{fig:gog_parallax}, right panels).

\section{Parallax zero-point}\label{sec:zero_point}

\subsection{Background}\label{subsec:zero_point_background}

    The global parallax zero-point offset in the {\it Gaia} DR2 data set should have come as no surprise. It has been known from {\it Hipparcos} times \cite[e.g.,][]{1995A&A...304...52A,1998A&A...340..309M} that a scanning, global space astrometry mission with a design such as that of {\it Hipparcos} and {\it Gaia} will have a (almost) full degeneracy between spin-synchronous variations of the basic angle between the (viewing directions of the) two telescopes and the zero-point of the parallaxes in the catalogue \citep[for details, see][]{2017A&A...603A..45B}. The zero-point offset was determined during the data processing and was published in \cite{2018A&A...616A...2L} as $-29 \pm 1~\mu$as, in the sense of {\it Gaia} parallaxes being too small, based on the median parallax of a sample of half a million primarily faint quasars contained in {\it Gaia} DR2. During the internal validation of the data processing prior to release, the zero-point was investigated using $\sim$30 different methods and samples, systematically resulting in a negative offset of order a few dozen $\mu$as \citep[see Table~1 in][]{2018A&A...616A..17A}. During these early inspections, hints already appeared that the zero-point offset depends on sky position, magnitude, and colour of the source.

Subsequent external studies, using a variety of methods and primarily bright stellar samples, and often combined with external data, often resulted in a zero-point offset of about $-50~\mu$as. Some recent examples, ordered from low to high offset values, include
$-28 \pm  2~\mu$as from \cite{2019MNRAS.tmp.2241S} based on mixture modelling of globular clusters,
$-31 \pm 11~\mu$as from \cite{2019ApJ...872...85G} based on eclipsing binaries,
$-35 \pm 16~\mu$as from \cite{2018MNRAS.481L.125S} based on asteroseismology,
$-41 \pm 10~\mu$as from \cite{2019MNRAS.486.3569H} based on asteroseismology,
$-42 \pm 13~\mu$as from \cite{2019AJ....158..105L} based on RR Lyraes,
$-46 \pm 13~\mu$as from \cite{2018ApJ...861..126R} based on classical Cepheids,
$-48 \pm  1~\mu$as from \cite{2020MNRAS.493.4367C} based on hierarchical modelling of red clump stars,
$-49 \pm 18~\mu$as from \cite{2018A&A...619A...8G} based on classical Cepheids,
$-50 \pm  5~\mu$as from \cite{2019gaia.confE..13K} based on asteroseismology,
$-52 \pm  2~\mu$as from \cite{2019MNRAS.489.2} based on APOGEE spectrophotometric distances,
$-53 \pm  9~\mu$as from \cite{2019ApJ...878..136Z} based on asteroseismology and spectroscopy,
$-54 \pm  6~\mu$as from \cite{2019MNRAS.487.3568S} based on {\it Gaia} DR2 radial velocities,
$-57 \pm  3~\mu$as from \cite{2018MNRAS.481.1195M} based on RR Lyraes,
$-75 \pm 29~\mu$as from \cite{2019ApJ...875..114X} based on VLBI astrometry,
$-76 \pm 25~\mu$as from \cite{2019arXiv190609827L} based on VLBI data of radio stars, and
$-82 \pm 33~\mu$as from \cite{2018ApJ...862...61S} based on eclipsing binaries.

\begin{table*}[th]
\caption{Frequency of occurrence of the first significant digit of the {\it Gaia} DR2 parallaxes (first line), a Lorentzian distribution with half-width $\gamma = 360~\mu$as (second line), a Lorentzian distribution with half-width $\gamma = 360~\mu$as and shifted by $+260~\mu$as (third line), and Benford's law (fourth line).\label{tab:lorentzian}}
\begin{center}
\begin{tabular}[h]{cccccccccl}
\hline\hline
\\[-8pt]
1 & 2 & 3 & 4 & 5 & 6 & 7 & 8 & 9 & Data / function\\
\hline
\\[-8pt]
0.256 & 0.167 & 0.140 & 0.116 & 0.093 & 0.075 & 0.061 & 0.051 & 0.043 & {\it Gaia} DR2 parallaxes\\
0.289 & 0.181 & 0.132 & 0.102 & 0.081 & 0.067 & 0.056 & 0.049 & 0.044 & Lorentzian\\
0.260 & 0.174 & 0.140 & 0.112 & 0.090 & 0.072 & 0.059 & 0.049 & 0.043 & Lorentzian shifted by $+260~\mu$as\\
0.301 & 0.176 & 0.125 & 0.097 & 0.079 & 0.067 & 0.058 & 0.051 & 0.046 & Benford's law\\
\hline
\end{tabular}
\end{center}
\end{table*}

\subsection{Varying the parallax zero-point offset}\label{subsec:zero_point_analysis}

\begin{figure}[t]
  \begin{center}
    \includegraphics[width=1.0\columnwidth]{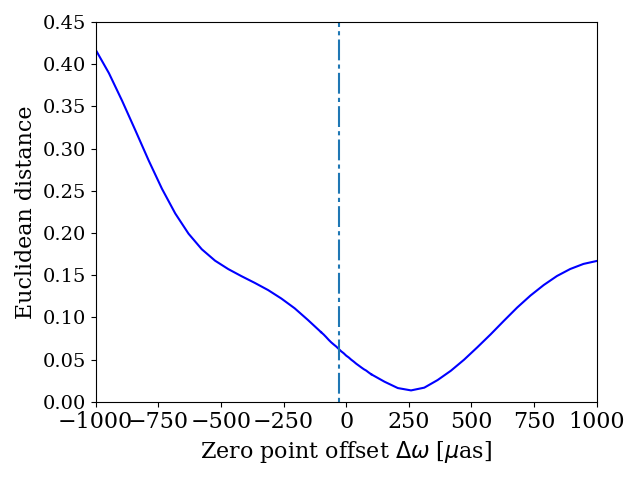}
    \caption{Euclidean distance between the distribution of the first significant digit of the {\it Gaia} DR2 parallaxes (see Appendix \ref{sec:stats}), after subtracting a trial zero-point offset $\Delta\varpi$ such that $\varpi_{\rm corrected} = \varpi_{Gaia~DR2} - \Delta\varpi$, and Benford's law for trial zero-point offsets $\Delta\varpi$ between $-1000$ and $1000~\mu$as. The dashed vertical line denotes the (faint-) QSO-based offset of $-29~\mu$as derived in \cite{2018A&A...616A...2L}, with more recent work suggesting that the relevant value for (bright) stars is about $-50~\mu$as (see Section~\ref{subsec:zero_point_background}).
    \label{fig:gaia_dr2_parallax_zero_point}}
  \end{center}
\end{figure}

The question arises whether the already fair agreement between the {\it Gaia} DR2 parallaxes and Benford's law, as discussed in Section~\ref{subsec:gaia_dr2_parallax} and displayed in Figure~\ref{fig:gaia_dr2_parallax}, would further improve when due account of the parallax zero-point offset would be taken. Naively, we would expect that a distribution of a quantity such as the parallax that covers several orders of magnitude, a small, uniform shift would not drastically change its behaviour with respect to Benford's law.

Our findings are summarised in Figure~\ref{fig:gaia_dr2_parallax_zero_point}. It shows for a range of trial zero-point offsets $\Delta\varpi$ the Euclidean distance between the distribution of the first significant digit of the {\it Gaia} DR2 parallaxes, after subtracting a zero-point offset $\Delta\varpi$ such that $\varpi_{\rm corrected} = \varpi_{Gaia~DR2} - \Delta\varpi$, and Benford's law. With this convention, the zero-point offset from \cite{2018A&A...616A...2L} translates into $\Delta\varpi = -29~\mu$as (see Section~\ref{subsec:zero_point_background}). The plot shows that changing the offset from $\Delta\varpi = 0$ to $-29~\mu$as only changes the Euclidean distance metric by 0.007, and even in the direction of worsening the agreement between the offset-corrected parallaxes and Benford's law.

A striking feature in Figure~\ref{fig:gaia_dr2_parallax_zero_point} is the pronounced minimum seen around $\Delta\varpi \sim +260~\mu$as. This minimum can be understood as follows. The {\it Gaia} DR2 parallax histogram itself (Figure~\ref{fig:gaia_dr2_parallax}) roughly resembles a Lorentzian of half-width $\gamma = 360~\mu$as and with a mean that is offset by some $-260~\mu$as. Table~\ref{tab:lorentzian} shows that such a Lorentzian has a distribution of first significant digits that already resembles that of Benford's law (see Appendix~\ref{subsec:lorentzian}), with digit 1 appearing most frequently and digit 9 appearing least frequently. By applying a uniform offset of $+260~\mu$as, the corrected parallax distribution becomes roughly symmetric and the match between the Lorentzian and the shifted {\it Gaia} DR2 data improves even further. We conclude that the conspicuous minimum in Figure~\ref{fig:gaia_dr2_parallax_zero_point} around $\Delta\varpi \sim +260~\mu$as has a mathematical reason, namely that this particular parallax offset causes the distribution of the shifted {\it Gaia} DR2 parallaxes to become optimally symmetric, instead of being caused by the zero-point offset.

\section{Summary and conclusions}\label{sec:conclusions}

We investigated whether Benford's law applies to {\it Gaia} DR2 data. Although it has been known for a long time that this law applies to a wide variety of physical data sets, it was only recently shown by \cite{2014JApA...35..639A} that it also holds for {\it Hipparcos} astrometry. We showed that the 1.3 billion observed parallaxes in {\it Gaia} DR2 follow Benford's law even closer. Stars with a parallax starting with digit $1$ are five times more numerous than stars with a parallax starting with digit $9$.

We reached a very different conclusion concerning the astrometric \emph{\textup{distance estimates}}. Using {\it Hipparcos} astrometry, \cite{2014JApA...35..639A} computed distance estimates as the reciprocal of the parallax, and found that this data set also follows Benford's law closely. However, \cite{2015PASP..127..994B} and \cite{2018A&A...616A...9L} showed that the reciprocal of the observed parallax $\varpi^{-1}$ is a poor estimate of the distance when the relative parallax error exceeds $\sim$10-20\%. The distance estimate can be improved by adding prior information about our Galaxy \citep{2015PASP..127..994B} and/or by including additional data such as photometry \citep{2019arXiv190411302A}. We unambiguously demonstrated that in neither case does the improved distance estimates follow Benford's law, although distances with small starting digits are still more abundant. Moreover, using realistic simulations of the stellar content of the Milky Way \citep{2012A&A...543A.100R}, we showed that the distances \emph{\textup{ought not}} to follow Benford's law, essentially because the interplay between the luminosity function of the Milky Way and {\it Gaia} mission selection function results in a bi-modal distance distribution, corresponding to nearby dwarfs in the Galactic disc and distant giants in the Galactic bulge. The fact that the true distances underlying the {\it Gaia} catalogue do not follow Benford's law, while the observed parallaxes do follow this law, probably due to observational errors, is the most intriguing result of this paper.

One of our objectives was to use Benford's law (or the deviation from it) as an indicator of anomalous behaviour, not necessarily giving hard evidence, but rather providing an indicator whose subsets warrant a deeper analysis \citep[e.g.][investigating money laundering]{moneylaundering}. We investigated the application of several astrometric and photometric quality filters applied to the {\it Gaia} DR2 parallaxes, but none changed the adherence to Benford's law by more than a few percent points.

Finally, we analysed the parallax zero-point that would be needed to optimise the fit to Benford's law, to compare it with the roughly $-50~\mu$as zero-point offset that is known to be present in the {\it Gaia} DR2 parallaxes \cite[e.g.][]{2019gaia.confE..13K}. An offset value of $+260~\mu$as was recovered. This can be understood by the negative tail of the Lorentzian-like {\it Gaia} DR2 parallax distribution, which for this offset value results in an optimally symmetric corrected-parallax distribution that closely follows Benford's law. We therefore conclude that Benford's law should not be used to validate the parallax zero-point in {\it Gaia} DR2.

\begin{acknowledgements}

  This work has made use of data from the European Space Agency (ESA) mission {\it Gaia} (\url{https://www.cosmos.esa.int/gaia}), processed by the {\it Gaia} Data Processing and Analysis Consortium (DPAC, \url{https://www.cosmos.esa.int/web/gaia/dpac/consortium}). Funding for the DPAC has been provided by national institutions, in particular the institutions participating in the {\it Gaia} Multilateral Agreement. We would like to thank Timo Prusti, Daniel Michalik, Alice Zocchi, and Eero Vaher for stimulating discussions and Eduard Masana for tips on how to create a (pseudo-)random query on the Paris-Meudon TAP server containing observed GUMS parallaxes. We would like to thank the anonymous referee for her/his constructive feedback. This research is based on data obtained by ESA’s {\it Hipparcos} satellite and has made use of the ADS (NASA), SIMBAD / VizieR (CDS), and TOPCAT \citep[][\url{http://www.starlink.ac.uk/topcat/}]{2005ASPC..347...29T} tools. We acknowledge support from the ESTEC Faculty Visiting Scientist Programme. The research leading to these results has received funding from the BELgian federal Science Policy Office (BELSPO) through PRODEX grants {\it Gaia} and {\it PLATO}.

\end{acknowledgements}

\bibliographystyle{BL} 
\bibliography{BL} 

\appendix

\section{Effect of {\it Gaia} DR2 quality filters}\label{sec:filters}

\begin{figure*}[t]
  \begin{center}
    \includegraphics[width=1.0\textwidth]{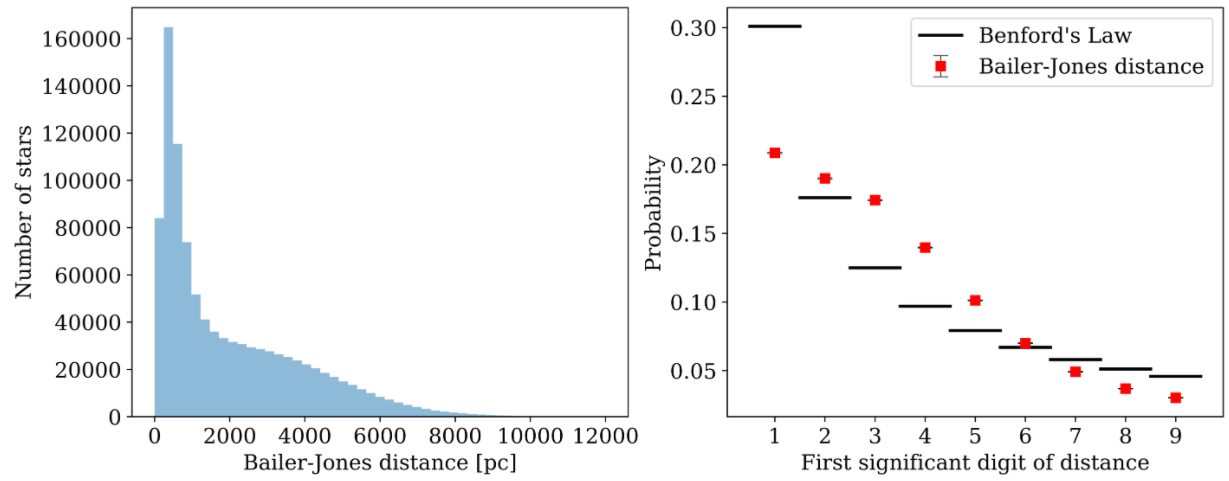}
    \caption{{\it Left:} Histogram of the {\it Gaia} DR2 Bayesian distance estimates from \cite{2018AJ....156...58B} for the sample of one million objects with the highest RUWE values, i.e. the objects with the poorest astrometric quality. {\it Right}: Distribution of the first significant digit of the Bayesian distance estimates, together with the theoretical prediction of Benford's law. The data have vertical error bars to reflect Poisson statistics, but the error bars are much smaller than the symbol sizes. Compare with Figure~\ref{fig:gaia_dr2_distance} for a sample of one million random stars that meet all astrometric (and photometric) quality criteria.
    \label{fig:gaia_dr2_distance_largest_RUWE}}
  \end{center}
\end{figure*}

\cite{2018A&A...616A...2L}, \cite{2018A&A...616A...4E}, and \cite{2018A&A...616A..17A}, all of whom were published together with and at the same date as the {\it Gaia} DR2 catalogue (25 April 2018), advocated using quality filters to define clean {\it Gaia} DR2 samples that are not hindered by astrometric and/or photometric artefacts. Such artefacts are known to be present in the data in particular in dense regions, and reflect the iterative and non-final nature of the data-processing strategy and status underlying {\it Gaia} DR2 and can be linked to erroneous observation-to-source matches, background subtraction errors, uncorrected source blends, etc. In this study, we employed the photometric excess factor filter as well as the astrometric quality filter, which is based on the renormalised unit weight error (RUWE) published post {\it Gaia} DR2 (in August 2018), requiring that valid sources meet the following two conditions:
\begin{equation}
  1.0 + 0.015 C^2 < E < 1.3 + 0.06 C^2,\label{eq:E}
\end{equation}
and
\begin{equation}
  {\rm RUWE} = \frac{\sqrt{\chi^2/(N-5)}}{u_0(G,C)} < 1.4,\label{eq:RUWE}
\end{equation}
where in the notation of the {\it Gaia} DR2 data model, $E = {\tt phot\_bp\_rp\_excess\_factor} = ({\tt phot\_bp\_mean\_flux} + {\tt phot\_rp\_mean\_flux}) / {\tt phot\_g\_mean\_flux}$, $C = {\tt bp\_rp} = {\tt phot\_bp\_mean\_mag} - {\tt phot\_rp\_mean\_mag}$, $\chi^2 = {\tt astrometric\_chi2\_al}$, $N = {\tt astrometric\_n\_good\_obs\_al}$, $G = {\tt phot\_g\_mean\_mag}$, and $u_0(G,C)$ is a look-up table as function of $G$ magnitude and $BP-RP$ colour index that is provided on the {\it Gaia} DR2 known issues webpage\footnote{
\url{https://www.cosmos.esa.int/web/gaia/dr2-known-issues}
}. These filters combined remove 620\,842\,302 entries from the {\it Gaia} DR2 catalogue (corresponding to 47\% of the data). The distribution of the first significant digit of the parallaxes is affected by application of the filters, but not to the extent that overall trends change. The maximum difference occurs for the frequency of digit 1, which equals 0.28 without filtering and 0.26 with the filtering applied.

Interestingly, and as a side note, the Bayesian distance estimates and associated first significant distribution of the sample of one million objects with the poorest astrometric quality (i.e. the highest RUWE values), displayed in Figure~\ref{fig:gaia_dr2_distance_largest_RUWE}, differ substantially from those derived from a random sample of filtered stars, as displayed in Figure~\ref{fig:gaia_dr2_distance}. With the evidence provided in the {\it Gaia} DR2 documentation and in the references quoted above that the astrometric quality filter is effective in removing genuinely bad and suspect entries, this is no surprise.

\section{Negative parallaxes}\label{sec:negative_parallaxes}

\begin{figure}[ht]
  \begin{center}
    \includegraphics[width=1.0\columnwidth]{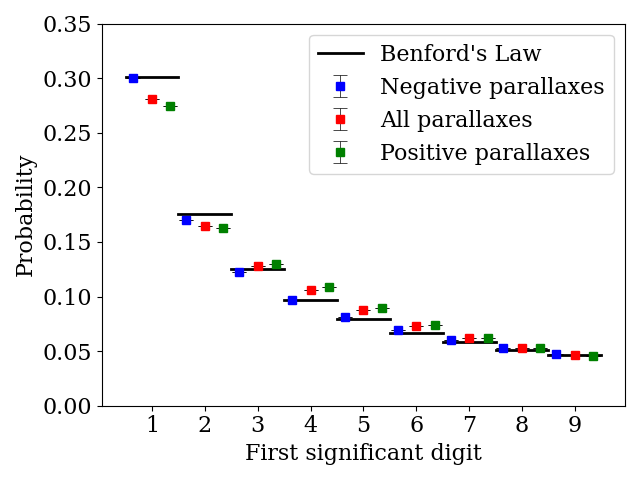}
    \caption{Distribution of the first significant digit of the {\it Gaia} DR2 parallaxes together with the theoretical prediction of Benford's law. Blue points refer to negative parallaxes (for which the statistics is based on the absolute value of the parallaxes), green points refer to positive parallaxes, and red points refer to the absolute value of all parallaxes (the red data can hence be considered as a weighted mean of the blue and green data with weights 18\% and 82\%, respectively).
    The data have vertical error bars to reflect Poisson statistics, but the error bars are much smaller than the symbol sizes.}
    \label{fig:gaia_dr2_negative_parallax}
  \end{center}
\end{figure}

Figure~\ref{fig:gaia_dr2_negative_parallax} shows that a significant fraction of the {\it Gaia} DR2 parallaxes is negative (26\% of published data, which reduces to 18\% when the filters discussed in Appendix~\ref{sec:filters} are applied). The complication of this fact is that there is no rule of how to deal with these data in relation to Benford's law. In addition, it is known that negative parallaxes are a totally normal and expected outcome of the astrometric measurement concept of {\it Gaia} and that simply ignoring negative parallaxes can lead to severe biases in the astrophysical interpretation of the data \cite[for a detailed discussion, see][]{2018A&A...616A...9L}. Figure~\ref{fig:gaia_dr2_negative_parallax} shows how sensitive the frequency distribution of the first significant digits is to the way in which we treat negative parallaxes: Variations in the frequency of occurrence of a few percent points appear depending on whether we include the negative parallaxes (after taking the absolute value) or not. Throughout this work, when we studied {\it Gaia} DR2 parallaxes and first-significant-digit statistics, we included all parallaxes, that is, we considered the absolute value of the measured parallax. This refers to the red symbols in Figure~\ref{fig:gaia_dr2_negative_parallax}, which can be interpreted as a reasonable (weighted-average) compromise between the two limiting cases defined by objects with positive parallax on the one hand (green symbols) and objects with negative parallax on the other hand (blue symbols).

\section{Tests with smaller sample sizes}\label{sec:sample_size}

\begin{table*}[th]
\caption{Frequency of occurrence of the first significant digit of the parallaxes in {\it Gaia} DR2 using randomly selected entries for various sample sizes. The column {\it Gaia} DR2 refers to $1\,332$M. K stands for $1\,000,$ while M stands for $1\,000\,000$.}\label{tab:sample_size}
\begin{center}
\begin{tabular}[h]{clllllll}
\hline\hline
\\[-8pt]
First significant digit & 1K & 10K & 100K & 1M & 10M & 100M & {\it Gaia} DR2\\
\hline
\\[-8pt]
1 & 0.292  & 0.284  & 0.281  & 0.281  & 0.281  & 0.281  & 0.281\\
2 & 0.160  & 0.159  & 0.167  & 0.164  & 0.164  & 0.164  & 0.164\\
3 & 0.122  & 0.125  & 0.127  & 0.128  & 0.128  & 0.128  & 0.128\\
4 & 0.098  & 0.109  & 0.105  & 0.106  & 0.106  & 0.106  & 0.106\\
5 & 0.106  & 0.0877 & 0.0870 & 0.0874 & 0.0876 & 0.0876 & 0.0876\\
6 & 0.0740 & 0.0701 & 0.0734 & 0.0728 & 0.0729 & 0.0729 & 0.0729\\
7 & 0.0671 & 0.0627 & 0.0620 & 0.0616 & 0.0617 & 0.0617 & 0.0617\\
8 & 0.0468 & 0.0548 & 0.0514 & 0.0529 & 0.0528 & 0.0528 & 0.0528\\
9 & 0.0356 & 0.0496 & 0.0463 & 0.0465 & 0.0460 & 0.0460 & 0.0460\\
\hline
\end{tabular}
\end{center}
\end{table*}

In view of the significant number of objects contained in {\it Gaia} DR2 and the associated non-negligible processing loads and run times, we conducted experiments to verify to what extent reduced sample sizes with randomly\footnote{
  We used the {\tt random\_index} field in {\it Gaia} DR2; see \url{https://gea.esac.esa.int/archive/documentation/GDR2/Gaia_archive/chap_datamodel/sec_dm_main_tables/ssec_dm_gaia_source.html}.
} selected objects return reliable results on the frequency distribution of the first significant digit of the parallaxes. Table~\ref{tab:sample_size} summarises our findings. It shows that percent-level accurate data can be derived from randomly selected samples of about one million objects. When we refer to {\it Gaia} DR2 statistics here, we consistently used samples containing one million randomly selected entries without inducing loss of generality.

\section{Statistics and Benford's law: justification of the Euclidean distance measure}\label{sec:stats}

We used the Euclidean distance to quantify how well the distribution of the first significant digit of {\it Gaia} data sets is described by Benford's law,
\begin{equation}
{\rm ED} = \sqrt{\Sigma_{d=1}^{9} \left(p_d - e_d\right)^2},\label{eq:ed}
\end{equation}
where $p_d$ is the measured digit frequency and $e_d$ is the expected digit frequency for digit $d$ according to Benford's law. The Euclidean distance ranges between 0 (when all first significant digits exactly follow Benford's law) and $1.036$ (when all first significant digits equal 9), with lower values indicating better adherence to Benford’s Law. Although the Euclidean distance is not a formal test statistic (see the discussion below), it is independent of sample size, which makes it suitable as a relative metric. The problem with sample-size dependent tests such as $\chi^{2}$ (or Kolmogorov-Smirnov) is evident from their definition,
\begin{equation}\label{eq:chi}
\chi^{2} = \sum_{d=1}^{9}\left(\frac{O_{d}-E_{d}}{\sigma}\right)^{2},   
\end{equation}
where $O_{d}$ is the observed number of occurrences of digit $d$, $E_{d}$ is the expected number of occurrences of digit $d$, and $\sigma$ reflects the measurement error on $O_{d}$,
\begin{eqnarray}
O_{d} &=& N p_{d}\nonumber\\
E_{d} &=& N e_{d} = N\log_{10}\left(1+\frac{1}{d}\right),
\end{eqnarray}
where $N$ is the total number of data points (stars in {\it Gaia} DR2 in our case, so $N \sim 10^9$). With counting (Poisson) statistics, $\sigma \propto \sqrt{N}$, such that $\chi^2 \propto N$. In practice, this means that a formal test statistic based on $\chi^2$ would reject many cases in which the data under test come from distributions that follow Benford's law \citep[e.g.][]{TamChoGaines}. In other words, because $\chi^2$ tests have large statistical power for high values of $N$, even quite small differences will be statistically significant, and because {\it Gaia} DR2 contains a billion parallaxes that are distributed over just nine bins ($d=1,\ldots,9$), the formal error bars on the observed frequencies in each bin (resulting just from Poisson statistics) are so tiny that any peculiarity in the data (e.g. an open cluster at a specific distance or  an extragalactic objects), would affect the statistical test and might provide misleading conclusions (exclusively looking at statistical errors, compliance of {\it Gaia} DR2 parallaxes with Benford’s Law would be excluded right away with very high significance levels). A reduced-$\chi^2$ statistic would also not alleviate this problem because the reduction would only divide $\chi^2$ by the number of bins (nine) and not by the number of data points ($N$). The Euclidean distance employed in this work, on the other hand, is independent of sample size and hence provides a metric that only becomes more precise with increasing sample size, but does not run away.

Clearly, a disadvantage of using the Euclidean distance is that it is not a formal test statistic with associated statistical power (although \citet{GoodmanED} suggested that data can be said to follow Benford's law when the Euclidean distance is shorter than $\sim$0.25). Many researchers have investigated and have proposed suitable metrics that can quantify statistical (dis)agreement between data and Benford's law (e.g. the Cram\'{e}r-von Mises metric; \citealt{Lesperance}). We did not explore such metrics further for several reasons that are essentially all linked to the existing freedom and arbitrariness in the interpretation of the {\it Gaia} DR2 data, as listed below. Note here that we mean that a 5\% increase of 0.4 (40\%) results in 0.42 (42\%), while a 5\%\ {\it \textup{point}} increase of 0.4 (40\%) results in 0.45 (45\%).
\begin{enumerate}
\item It is known that {\it Gaia} DR2 contains in addition to a small fraction of non-filtered duplicate sources (cf.\ Figure~2 in \citealt{2018A&A...616A..17A}) genuine sources with spurious astrometry (and/or photometry). As explained in Appendix~\ref{sec:filters}, several filters have been recommended to obtain clean data sets (e.g. RUWE, astrometric excess noise, photometric excess factor, number of visibility periods, and the longest semi-major axis of the five-dimensional astrometric error ellipsoid; \citealt{2018A&A...616A...2L}). In all of these cases, however, the specific threshold values to be used, and also which combination of filters to be used, is specific to the science application, without an absolute truth. Depending on subjective choices, the observed distribution over the first significant digits changes by up to several percent points (see Appendix~\ref{sec:filters}), which is orders of magnitude larger than the formal statistical errors (one~billion stars divided equally over nine bins corresponds to a Poisson error of about $\sqrt{10^{-8}} = 10^{-4}$ per bin).
\item There is ambiguity on how negative parallaxes (comprising 26\% of the published {\it Gaia} DR2 data), which are perfectly valid measurements, should be treated. Again, depending on subjective choices, the observed distribution over the first significant digits changes by up to several percent points (see Appendix~\ref{sec:negative_parallaxes} and Figure~\ref{fig:gaia_dr2_negative_parallax}).
\item In the same spirit, arbitrary changes of units (e.g. from parsec to light years [1 pc = 3.26 ly] or milliarcseconds to nanoradians [1 mas = 4.85 nrad]) change the observed distribution over the first significant digits by up to several percent points.
\item It is known that the {\it Gaia} DR2 parallaxes collectively have a global parallax zero-point offset, which to second order depends on magnitude, colour, and sky position (see Section~\ref{sec:zero_point} for a detailed discussion). Again, the absolute truth is out there, and depending on the subjective choice for the value of the offset correction, the observed distribution over the first significant digits changes significantly (see Figure~\ref{fig:gaia_dr2_parallax_zero_point}).
\end{enumerate}
In short, even with a proper statistical test or metric, we would not be able to capture the effects of the existing freedom (and systematic effects) in the (interpretation of the) data.

\onecolumn

\section{ADQL queries}\label{sec:adql}

The following ADQL queries, and slight variations thereof, were used in this research:
\begin{itemize}
\item
To query {\it Gaia} DR2 data from the {\it Gaia} Archive at ESA (\url{https://gea.esac.esa.int/archive/}):
{\null\vskip-0.5truecm\null
\begin{verbatim}
SELECT BailerJones.source_id, BailerJones.r_est, BailerJones.r_len, Gaia.parallax,
Gaia.parallax_over_error FROM external.gaiadr2_geometric_distance as BailerJones INNER JOIN
(SELECT GaiaData.source_id, GaiaData.parallax, GaiaRUWE.ruwe FROM gaiadr2.gaia_source as
GaiaData INNER JOIN (SELECT * FROM gaiadr2.ruwe where ruwe < 1.4) as GaiaRUWE ON
GaiaData.source_id = GaiaRUWE.source_id where (GaiaData.phot_bp_rp_excess_factor < 
1.3 + 0.06 * POWER(GaiaData.phot_bp_mean_mag - GaiaData.phot_rp_mean_mag, 2) and
GaiaData.phot_bp_rp_excess_factor > 1 + 0.015 * POWER(GaiaData.phot_bp_mean_mag -
GaiaData.phot_rp_mean_mag, 2))) as Gaia ON BailerJones.source_id = Gaia.source_id;
\end{verbatim}
\null\vskip-0.75truecm\null
}
\item
To query median StarHorse distance estimates for a random subset of high-quality objects from the {\it Gaia} Archive at the Leibniz Institute for Astrophysics in Potsdam (\url{https://gaia.aip.de/query/}; see Section~\ref{subsec:gaia_dr2_distance}):
{\null\vskip-0.5truecm\null
\begin{verbatim}
SELECT TOP 1000000 StarHorse.source_id, StarHorse.dist50, Gaia.parallax FROM
gdr2_contrib.starhorse as StarHorse INNER JOIN (SELECT source_id, parallax FROM
gdr2.gaia_source ORDER BY random_index) as Gaia ON StarHorse.source_id =
Gaia.source_id AND StarHorse.SH_OUTFLAG LIKE '00000' AND StarHorse.SH_GAIAFLAG LIKE '000'
\end{verbatim}
\null\vskip-0.75truecm\null
}
\item
To query true GUMS distances from the {\it Gaia} Archive at the Centre de Donn\'ees astronomiques de Strasbourg (\url{http://tapvizier.u-strasbg.fr/adql/?gaia}; \citealt{2000A&AS..143...23O}; see Section~\ref{subsec:simulations}):
{\null\vskip-0.5truecm\null
\begin{verbatim}
SELECT "VI/137/gum_mw".r FROM "VI/137/gum_mw",
\end{verbatim}
after which a random sample of one million objects was selected using the shuf command.
\null\vskip-0.25truecm\null
}
\item
To query one million random observed GUMS parallaxes from the {\it Gaia} Archive at the Observatoire de Paris-Meudon (\url{https://gaia.obspm.fr/tap-server/tap}; see Section~\ref{subsec:simulations}):
{\null\vskip-0.5truecm\null
\begin{verbatim}
SELECT parallax FROM simus.complete_source,
\end{verbatim}
after which a random sample of one million objects was selected using the shuf command.
\null\vskip-0.5truecm\null
}
\end{itemize}

\section{Selected mathematical derivations}\label{sec:BL_details}

For convenience of the reader, without pretending to have derived these relations as new discoveries, this appendix presents selected derivations linked to scale invariance (Appendix~\ref{subsec:scale_invariance}), base invariance (Appendix~\ref{subsec:base_invariance}), and inverse invariance (Appendix~\ref{subsec:inverse_invariance}; see e.g. \citealt{MR1233974} and \citealt{wolfram_mathworld}). Appendix~\ref{subsec:lorentzian} discusses the distribution of first significant digits of a Lorentzian distribution.

\subsection{Scale invariance}\label{subsec:scale_invariance}

It is possible to define the probability for the first significant digit with a probability density function as follows:
\begin{equation}\label{eq:E11}
P(D_{1}X=d) = P(\lfloor{x}\rfloor=d) = P(d\leq x<d+1) = \int_{d}^{d+1}p(x)\text{d}x.
\end{equation}
If Benford's law is a universal law, it needs to be independent of the selected unit (e.g. parsec or light year). In other words, the first significant digit distribution has to be scale invariant. If the distribution of data set $\tilde{X}$ is scale invariant, then there exists $\forall X \in \tilde{X}$, a scale $C \in \mathbb{R}_{> 0}$, an $\alpha \in (0,10)$, and a function $f$ such that for a significand $x$ of $X$, we have
\begin{equation}\label{eq:E13}
P\left(D_{1}(C X)=d\right) = P(\lfloor CX \rfloor=d) = P(\lfloor\alpha x\rfloor=d) = P(\lfloor x\rfloor=d)/f(\alpha),
\end{equation}
where
\begin{eqnarray} 
 \alpha &=& \frac{c}{10}\text{\quad if the significand of $CX$ is smaller than the significand of $X$;}\\
 \alpha &=& c \text{\quad if the significand of $CX$ is larger than or equal to the significand of $X$,}
\end{eqnarray}
such that
\begin{equation}
p(x)=f(\alpha)\cdot p(\alpha x).\label{eq:p}
\end{equation}
We exclude $\alpha=C=0$ because this is a special case for which Eq.~(\ref{eq:E13}) gives $d=0$, when $P\left(D_{1}(0\cdot X)=d\right)$ for every $X$.

In order to prove that Benford's law appears if and only if the data are scale invariant, we start with assuming that the data follow Benford's law. This means that the probability for the first significant digit $d$ of significand $x$ to appear equals
\begin{equation}
P(d\leq x<d+1) = \log_{10}\left(\frac{d+1}{d}\right) = \frac{1}{\ln\left(10\right)}\int_{d}^{d+1}\frac{1}{x}\text{d}x.
\end{equation}
Therefore the probability density function equals
\begin{equation}
p(x) = \frac{1}{\text{ln}(10)\cdot x}.
\end{equation}
This probability density function satisfies Eq.~(\ref{eq:p}) for $f(\alpha) = \alpha$. This proves that if a data set follows Benford's law, it is scale invariant.

Next, we assume that the data are scale invariant. This implies that Eq.(\ref{eq:p}) holds for every $\alpha$ 
$\in \mathbb{R}_{\geq 0}$. If $p(x)$ is a continuous probability density function on $[1,10)$, such that
\begin{equation}
\int_{1}^{10} p(x)\text{d}x=1,
\end{equation} 
we can derive
\begin{equation}
P(1\leq x<10) = \int_{1}^{10} p(\alpha x)\text{d}x = \int_{1}^{10}p(z)\frac{\text{d}z}{\alpha} = \frac{1}{\alpha},
\end{equation}
where $z\equiv\alpha x$. This result gives $f(\alpha) = \alpha$. Therefore
\begin{equation}
\frac{p(x)}{\alpha}=p(\alpha x).
\end{equation}
By taking the derivative of both sides with respect to $\alpha$, and by choosing $\alpha=1$, the following relation holds:
\begin{equation}
-p(x) = x\frac{\partial p(x)}{\partial x}.
\end{equation}
This differential equation can be solved with the separation technique and gives
\begin{eqnarray}
-\ln(x)+c&=&\ln(p(x));\nonumber\\
\frac{\lambda}{x}&=&p(x),
\end{eqnarray}
where $\lambda=e^{c}$. Now, it is possible to derive
\begin{equation}
\int_{1}^{10}p(x)\text{d}x = \int_{1}^{10}\frac{\lambda}{x}\text{d}x = \lambda\left[\ln(10)-\ln(1)\right]=1,
\end{equation}
so that
\begin{equation}
\lambda=\frac{1}{\ln\left(10\right)},
\end{equation}
and the probability density function $p(x)$ becomes
\begin{equation}
p(x)=\frac{1}{\ln\left(10\right)\cdot x}.
\end{equation}
The probability of the first significant digit can now be derived by
\begin{equation}
P(D_{1}X=d) = P(d\leq x<d+1) = \int_{d}^{d+1} p(x) \text{d}x = \frac{1}{\ln\left(10\right)}\int_{d}^{d+1}\frac{1}{x}\text{d}x = \log_{10}\left(\frac{d+1}{d}\right).
\end{equation}
This is exactly Benford's law for the first significant digit (Eq.~\ref{eq:BL}). This proves that if the data are scale invariant, they follow Benford's law.
In conclusion, a necessary precondition for a data set to follow Benford's law is scale invariance.

\subsection{Base invariance}\label{subsec:base_invariance}

If Benford's law is a universal law, it should be base invariant as well, next to being scale invariant, as these properties have a common origin.\footnote{As already mentioned in Section~\ref{sec:BL}, \citet{MR1421567} demonstrated that scale invariance implies base invariance (but base invariance does not imply scale invariance).} Consider for example the scale invariance of a uniform logarithmic distribution, which shows that base invariance is related to scale invariance. We can generalise the scale-invariance derivation in Appendix~\ref{subsec:scale_invariance} by substituting base 10 with base $B$ such that
\begin{equation}
\int_{1}^{B}p(x)\text{d}x = \int_{1}^{B}\frac{\lambda}{x}\text{d}x = \lambda\left[\ln(B)-\ln(1)\right] = 1.
\end{equation}
Therefore we find
\begin{equation}
\lambda=\frac{1}{\ln\left(B\right)},
\end{equation}
which gives
\begin{equation}
P(D_{1}X=d) = \int_{d}^{d+1} p(x) \text{d}x = \log_{B}\left(\frac{d+1}{d}\right).
\end{equation}
Base invariance was discussed in detail by \citet{MR1233974}.

\subsection{Inverse distribution}\label{subsec:inverse_invariance}

When parallaxes and distances are discussed, a relevant question is the relation between a data set $\tilde{X}$ and its inverse $\tilde{X}^{-1}$. From the scale invariance of a uniform logarithmic distribution, we might intuitively already expect that the inverse distribution is scale invariant as well. We here formally demonstrate that if $\tilde{X}$ is a data set that follows Benford's law, then the inverted data $\tilde{X}^{-1}$ also follow Benford's law.

First, we note that the mapping of the mantissae to the inverse of the mantissae is given by
\begin{equation}
\left\{\tilde{X}\rightarrow\tilde{X}^{-1} : x \mapsto x^{-1}\cdot 10^{b}\right\},
\end{equation}
where $b=0$ if $x=10^{n}$ $\forall n\in\mathbb{Z}$ and $b=1$ for any other value of $x$.

Next, we assume that $\tilde{X}$ follows Benford's law, such that for $\forall X \in \tilde{X}$ in significand notation $X = x \cdot 10^{m}$, with $m \in \mathbb{Z}$, we have
\begin{eqnarray}
  P(D_{1}X^{-1}=d) &=& P(x^{-1} \in [d,d+1)) = P(d \leq x^{-1} < d+1) = P\left(\frac{10^{b}}{d+1} < x \leq \frac{10^{b}}{d}\right) = \frac{1}{\ln(10)}\int_{\frac{10^{b}}{d+1}}^{\frac{10^{b}}{d}} \frac{1}{x}\text{d}x = \nonumber\\
&=& \frac{1}{\ln(10)}\int_{\frac{1}{d+1}}^{\frac{1}{d}} \frac{1}{x}\text{d}x = \frac{1}{\ln(10)}\left[-\ln(d)+\ln(d+1)\right] = \log_{10}\left(\frac{d+1}{d}\right).
\end{eqnarray}
This result, together with scale invariance (Appendix~\ref{subsec:scale_invariance}), implies that the mapping
$X \mapsto \alpha X^{-1}$
preserves Benford's law for any value $\alpha \in \mathbb{R}_{> 0}$.

\subsection{Distribution of first significant digits of a Lorentzian distribution}\label{subsec:lorentzian}

The normalised Lorentzian function, centred at $x=x_{0}$ and with a half width at half maximum of $\gamma$, is given by
\begin{equation}
 L(x; x_{0},\gamma)=\frac{1}{\pi\cdot\gamma\left(1+\left[\frac{x-x_{0}}{\gamma}\right]^{2}\right)}.\label{eq:Lorentzian}
\end{equation}
The first significant digit probability distribution for a Lorentzian function can be derived analytically by using the generic first significant digit probability function:
\begin{equation}
P(D_{1}X=d)=\sum_{k=-\infty}^{\infty}\int_{d\cdot 10^{k}}^{(d+1)\cdot 10^{k}}p(x)\text{d}x,
\end{equation}
and replacing the generic probability density function $p(x)$ with the Lorentzian from Eq.~(\ref{eq:Lorentzian}):
\begin{equation}
P(D_{1}X=d;x_{0},\gamma) =
\frac{\sum_{k=-\infty}^{\infty}\int_{d\cdot 10^{k}}^{(d+1)\cdot 10^{k}}L(x;x_{0},\gamma)\text{d} x}{\int_{0}^{\infty}L(x;x_{0},\gamma)\text{d} x} =
\sum_{k=-\infty}^{\infty}\left[\frac{\text{atan}\left(\frac{(d+1)\cdot 10^{k}-x_{0}}{\gamma}\right)-\text{atan}\left(\frac{d\cdot 10^{k}-x_{0}}{\gamma}\right)}{\frac{\pi}{2}+\text{atan}\left(\frac{x_{0}}{\gamma}\right)}\right].
\end{equation}
The normalisation assumes that only positive first significant digit values are considered ($x>0$).

With the atan addition formula, given by
\begin{equation}
\text{atan}(x)+\text{atan}(y)=
\begin{cases}
\text{atan}\left(\frac{x+y}{1-xy}\right) & \text{if $xy<1$},\\
\text{atan}\left(\frac{x+y}{1-xy}\right)+\pi & \text{if $xy>1$ and $x>0$ and $y>0$},\\
\text{atan}\left(\frac{x+y}{1-xy}\right)-\pi & \text{if $xy>1$ and $x<0$ and $y<0$},\label{eq:atan}
\end{cases}
\end{equation}
and with the following two identities (valid for $x>0$):
\begin{equation}
\text{atan}(-x)=-\text{atan}(x),
\end{equation}
\begin{equation}
\text{atan}\left(\frac{1}{x}\right)=\text{acot}(x),
\end{equation}
the first significant digit probability function can be written as
\begin{equation}
P(D_{1}X=d;x_{0},\gamma)= 
\sum_{k=-\infty}^{\infty}\left[\frac{\text{acot}\left(\gamma\cdot 10^{-k}+x_{0}^{2}\cdot 10^{-k}-(2d+1)x_{0}+d(d+1)10^{k}\right)}{\frac{\pi}{2}+\text{atan}\left(\frac{x_{0}}{\gamma}\right)}\right].
\end{equation}

In view of Eq.~(\ref{eq:atan}), a minor modification of the above result is required for index $k=k^{*}$, defined as
\begin{equation}
d\cdot 10^{k^{*}} < x_{0} < (d+1)\cdot 10^{k^{*}},
\end{equation}
and
\begin{equation}
\left((d+1)\cdot 10^{k^{*}}-x_{0}\right)\left(d\cdot 10^{k^{*}}-x_{0}\right)+\gamma^{2} < 0,
\end{equation}
such that the final result reads
\begin{eqnarray}\label{acot}
P(D_{1}X=d;x_{0},\gamma)
&=& 
\sum_{k=-\infty, k\neq k^{*}}^{\infty}\left[\frac{\text{acot}\left(\gamma\cdot 10^{-k}+x_{0}^{2}\cdot 10^{-k}-(2d+1)x_{0}+d(d+1)10^{k}\right)}{\frac{\pi}{2}+\text{atan}\left(\frac{x_{0}}{\gamma}\right)}\right] + \nonumber\\
&+& 
\sum_{k=k^{*}}\left[\frac{\text{acot}\left(\gamma\cdot 10^{-k}+x_{0}^{2}\cdot 10^{-k}-(2d+1)x_{0}+d(d+1)10^{k}\right)}{\frac{\pi}{2}+\text{atan}\left(\frac{x_{0}}{\gamma}\right)}+\pi\right].
\end{eqnarray}

\end{document}